\documentclass[runningheads]{llncs}
\usepackage{makecell}
\usepackage{graphicx}
\usepackage{amssymb}
\usepackage{framed,multirow,color}
\usepackage{marvosym}
\usepackage{threeparttable}
\begin{document}
\title{Is Dataset Quality Still a Concern in Diagnosis Using Large Foundation Model?}

\author{Ziqin Lin\inst{1,2*} \and 
Heng Li\inst{1*}\textsuperscript{\Letter} \and 
Zinan Li\inst{1,2} \and 
Huazhu Fu\inst{3} \and 
Jiang Liu\inst{1,4,5}\textsuperscript{\Letter}}
%
\institute{Research Institute of Trustworthy Autonomous Systems, Southern University of Science and Technology, Shenzhen, China\and
Department of Computer Science and Engineering, Southern University of Science and Technology, Shenzhen, China \and 
Institute of High Performance Computing, Agency for Science, Technology and Research, Singapore\and
Guangdong Provincial Key Laboratory of Brain-inspired Intelligent Computation, Southern University of Science and Technology, Shenzhen, China\and
Research Institute of Trustworthy Autonomous Systems, Southern University of Science and Technology, Shenzhen, China\and
Department of Electrical and Computer Engineering\\
\email{lih3, liuj@sustech.edu.cn}}


\maketitle              
\begin{abstract}
Recent advancements in pre-trained large foundation models (LFM) have yielded significant breakthroughs across various domains, including natural language processing and computer vision. 
These models have been particularly impactful in the domain of medical diagnostic tasks.
With abundant unlabeled data, an LFM has been developed for fundus images using the Vision Transformer (VIT) and a self-supervised learning framework. This LFM has shown promising performance in fundus disease diagnosis across multiple datasets. On the other hand, deep learning models have long been challenged by dataset quality issues, such as image quality and dataset bias.
To investigate the influence of data quality on LFM, we conducted explorations in two fundus diagnosis tasks using datasets of varying quality. 
Specifically, we explored the following questions: Is LFM more robust to image quality? Is LFM affected by dataset bias? Can fine-tuning techniques alleviate these effects?
Our investigation found that LFM exhibits greater resilience to dataset quality issues, including image quality and dataset bias, compared to typical convolutional networks. Furthermore, we discovered that overall fine-tuning is an effective adapter for LFM to mitigate the impact of dataset quality issues.

\end{abstract}

\section{Introduction}

Recent advancements in LFM have revolutionized the field of natural language processing (NLP) and computer vision (CV) as a whole.
These models, such as GPT-3 \cite{advances} and its successors, have achieved remarkable performance in various NLP tasks, including language generation, translation, summarization, and question-answering. 
In the CV region, Meta introduced the Segment Anything Model (SAM) \cite{SAM} as a versatile and scalable solution for image segmentation.
Moreover, LFM has shown remarkable potential in revolutionizing healthcare by analyzing medical text, assisting clinical decision-making, accelerating drug discovery, and contributing to public health surveillance \cite{cui2024scgpt,he2023accuracy,huang2024segment,li2023artificial,wu2023medical}. 
Therefore, due to the vast amount of fundus image data, Zhou et al. proposed RETFound \cite{RETFound}, a foundational model for retinal images, which
learns generalized representations from mass unlabeled retinal images and serves as a basis for efficient model adaptation to various applications with a few labeled data. Specifically, RETFound utilizes a modified version of  VIT \cite{VIT} as the base model, trained on 1.6 million unlabeled retinal images through self-supervised learning, followed by adaptation to disease detection tasks with a few explicit labels. RETFound consistently outperforms multiple comparative models in the diagnosis and prognosis of vision-threatening eye diseases, as well as event prediction for complex systemic diseases such as heart failure and myocardial infarction, despite having fewer labeled data.

On the other hand, dataset quality remains a common issue in clinical practice \cite{lin2023domain,SAME,SCR}, which can adversely affect both physicians’ observations and algorithmic diagnoses. The dataset quality not only encompasses image quality but also includes data bias. Data bias leads models to focus more on classes that are overrepresented in the dataset while neglecting classes with lower proportions. \textit{1) Can LFM withstand the influence of image quality?
2) Does dataset bias also impact the performance of LFM? 3) how do different fine-tuning techniques affect LFM?}

To explore the aforementioned three issues, we used RETFound as the representative foundation model and validated its performance on different quality subsets of the iSee and EyeQ of fundus datasets, respectively. We observed that RETFound is susceptible to a certain degree of influence from different datasets and images with varying qualities.
Moreover, we found that fine-tuning the parameters of the whole model can reduce the impact of image degradation.
Our main contributions are summarised as follows:
\begin{itemize}

    \item[{$\bullet$}] We validated the performance of RETFound on different quality subsets within the identical dataset, investigating how dataset quality affects LFM. Although the classification performance of both RETFound and ResNet decreases with the deterioration of the quality of the dataset, RETFound tends to be relatively stable.
    
    \item[{$\bullet$}] We validated the performance of RETFound on different categories within the identical dataset, investigating how dataset bias affects LFM. The existence of dataset bias can impact the performance of RETFound, but reducing dataset bias can mitigate the impact of image degradation on the performance of RETFound.
    
    \item[{$\bullet$}] We evaluated the performance of RETFound with different fine-tuning strategies on the identical dataset, exploring how fine-tuning alleviates the impact of dataset quality on RETFound. We evaluated three fine-tuning techniques: linear probe, updating the whole model, and Tip-Adapter \cite{tip}. As a result, updating the parameters of the whole model demonstrated outstanding performance.

\end{itemize}

\begin{figure}[!t]
    \begin{centering}
        \includegraphics[width=0.9\linewidth]{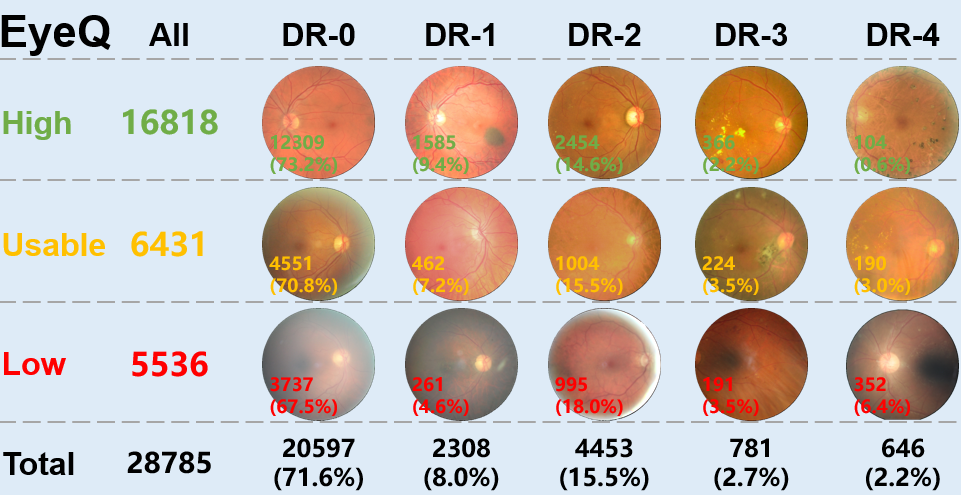}
        \par
    \end{centering}
\caption{Summary of our EyeQ dataset, where DR-i indicates the presence of diabetic retinopathy at grade i based on the labels in the EyePACS dataset. This image shows the three levels of quality of the EyeQ dataset (high, usable, and low) corresponding to each level of DR severity. Number (rate), where number represents the quantity of images in this category, and rate indicates the proportion of images of this DR category in the corresponding quality subset. 
}
\vskip -5pt
\label{fig:EyeQ_dataset}
\end{figure}

\section{Experiment Setting}
To delve into the impact of dataset quality and diverse tasks on LFM, it is imperative to select a dataset that encompasses both disease classification labels and quality grading labels. This step is pivotal in ensuring comprehensive exploration. Consequently, the emergence of the Eye-Quality (EyeQ) and Fundus-iSee (iSee) datasets addresses this crucial need, providing a robust foundation for in-depth investigations.

\begin{figure}[!t]
    \begin{centering}
        \includegraphics[width=0.9\linewidth]{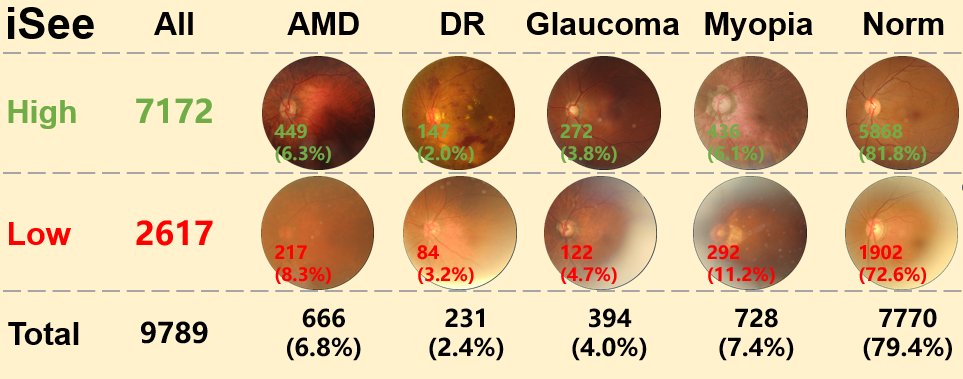}
        \par
    \end{centering}
\caption{Summary of our iSee dataset, which shows the fundus disease distribution. This image displays
the iSee dataset’s two levels of quality (high and low) correspond to each level of the fundus disease category. }
\vskip -5pt
\label{fig:iSee_dataset}
\end{figure}

\subsection{Dataset} 

\textbf{EyeQ:}
The EyeQ \cite{MCF} dataset comprises a total of 28,785 retinal images, categorized into three distinct quality levels: high-quality, usable quality, and low quality. Among these images, 16,818 are classified as high-quality, 6,431 as usable-quality, and 5,536 as low-quality. Each quality category is further delineated based on the severity levels of Diabetic Retinopathy (DR), denoted as DR-0, DR-1, DR-2, DR-3, and DR-4.

From the Fig. \ref{fig:EyeQ_dataset}, we can detailly observe the distribution of images within each quality category. The EyeQ dataset exhibits severe data imbalance, with DR-0 comprising the highest proportion, while DR-1, DR-3, and DR-4 are all represented by very low proportions.\\
\textbf{iSee:}
A fundus dataset comprises a total of 9789 images, with 7172 categorized as high-quality and 2617 as low-quality. These images are further classified into five categories: AMD (Age-related Macular Degeneration), DR, glaucoma, myopia, and norm (normal).

From the Fig. \ref{fig:iSee_dataset}, we can detailly observe the distribution of images within each quality category. The iSee dataset exhibits severe data imbalance, with norm comprising the highest proportion, while AMD, DR, glaucoma, and myopia are all represented by very low proportions.

\subsection{The implementation}
Here, due to the ability of the resblock to alleviate the problem of disapperance of the gradient, we utilized ResNet \cite{koonce2021resnet} as a comparative diagnostic classification model. We use RETFound \cite{RETFound} one of LFM as the diagnostic model, which is specified for Transformer-based networks \cite{attention}. Moreover, we use the Tip-Adapter \cite{tip} to fine-tune the RETFound as another base diagnosis model.

In our experimental setup, we partitioned high-quality subsets from the EyeQ and iSee datasets into training and testing sets at ratios of 0.8 and 0.2, respectively. The training set derived from the high-quality subset was employed for model training. Subsequently, the trained model underwent validation on subsets of varying quality levels from EyeQ (high, usable, and low) and iSee (high and low). We investigated the impact of dataset quality on the grading of DR (characterized by minimal differences) and the classification of fundus diseases (characterized by substantial differences).

Three deep learning models, ResNet, RETFound, and Tip-Adapter based on Contrastive Language-Image Pre-training (CLIP) \cite{clip}, were trained for 50 epochs with a learning rate of 0.001. The AdamW optimizer was employed for model optimization. Images were resized to 512x512 pixels before being fed into the models for processing. The classified performance was quantified by Area Under the Receiver Operating Characteristic curve(AUROC). To ensure the rigor of the experiments, we conducted each set of experiments five times and averaged the results.

\section{Analysis}

\subsection{The Impact of  Image Quality on RETFound}

\begin{figure}[!t]
    \begin{centering}
        \includegraphics[width=0.9\linewidth]{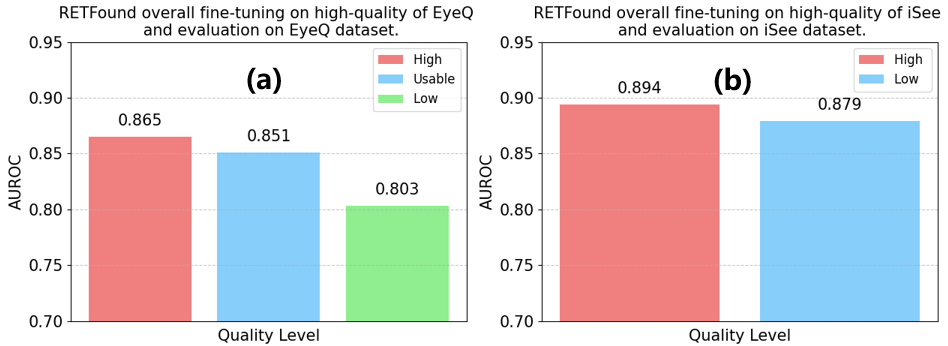}
        \par
    \end{centering}
\caption{RETFound overall fine-tuning on the high-quality EyeQ and then evaluated on low-quality, usable quality, and high-quality EyeQ datasets. (a) represents the impact of different qualities on RETFound in EyeQ, while (b) represents the impact of different qualities on RETFound in iSee. The classification performance of RETFound decreases as image quality deteriorates.
}
\vskip -5pt
\label{fig:REFTound on iSee and EyeQ}
\end{figure}

\begin{figure}[!t]
    \begin{centering}
        \includegraphics[width=0.9\linewidth]{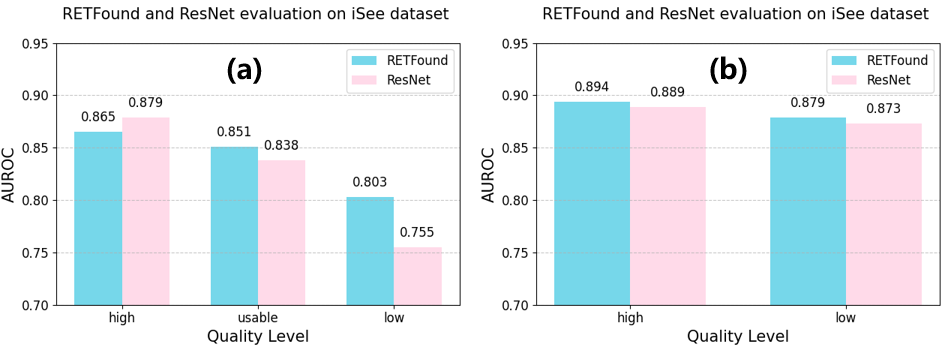}
        \par
    \end{centering}
\caption{RETFound and ResNet overall fine-tuning the training dataset of EyeQ and iSee high-quality subset, respectively. (a) and (b) show that RETFound is more stable than ResNet when encountering image degradation.
}
\vskip -5pt
\label{fig:RETFound_vs_Resnet}
\end{figure}


To investigate the impact of image quality on RETFound, we performed overall fine-tuning of RETFound on the high-quality training sets of EyeQ and iSee datasets, respectively, and evaluated its performance on their subsets. Detailed analysis and results are presented below:

\textbf{EyeQ: }From Fig. \ref{fig:REFTound on iSee and EyeQ} (a), in the evaluation of RETFound's performance in detecting DR grades across three subsets—EyeQ high quality (AUROC is 0.865), usable quality (AUROC is 0.851), and low quality (AUROC is 0.803).
The AUROC  of 0.865 indicates a good performance in predicting DR grades on the EyeQ high-quality dataset. On the usable-quality dataset, with an AUROC  of 0.851, the performance of the model slightly decreases compared to the high-quality dataset. The AUROC  falls further to 0.803 on the low-quality dataset, indicating a notable decline in performance.
This suggests that the model struggles significantly when processing low-quality images, resulting in reduced accuracy in predicting DR grades. The analysis demonstrates a clear correlation between the quality of the dataset and the performance of the model to detect the DR grades.

\textbf{iSee: }From Fig. \ref{fig:REFTound on iSee and EyeQ} (b), in evaluating RETFound's performance in detecting five categories of fundus diseases on the iSee dataset across two subsets—high-quality(AUROC is 0.894) and low-quality (AUROC is 0.879). With an AUROC  of 0.879, the model performance remains relatively high even when faced with images of lower quality.
Overall, the classification performance of RETFound is influenced by the quality of fundus images, which decreases as the image quality deteriorates.

To further explore the difference between RETFound and ResNet affected by image degradation, we also evaluated ResNet.
From Fig. \ref{fig:RETFound_vs_Resnet}, we observe that the performance of LFM and ResNet is significantly affected by image degradation, highlighting the critical importance of image quality enhancement techniques. Leveraging extensive parameterization and pre-training on massive datasets, LFM exhibits superior performance over ResNet in tasks such as DR grading and fundus disease classification (except for the high-quality subset of EyeQ ), with the degradation of image quality resulting in a smaller decline in classification performance for LFM compared to ResNet. While both RETFound and ResNet are trained on datasets comprising high-quality images, RETFound's pretraining spans datasets of variable quality, thus influencing its fine-tuning performance. Consequently, RETFound exhibits inferior performance compared to ResNet on the high-quality subset of EyeQ.

In summary, ResNet demonstrates pronounced sensitivity to image quality degradation, with performance exhibiting a substantial decline as image quality diminishes, while LFM exhibits greater robustness. 



\begin{figure}[!b]
    \begin{centering}
        \includegraphics[width=0.8\linewidth]{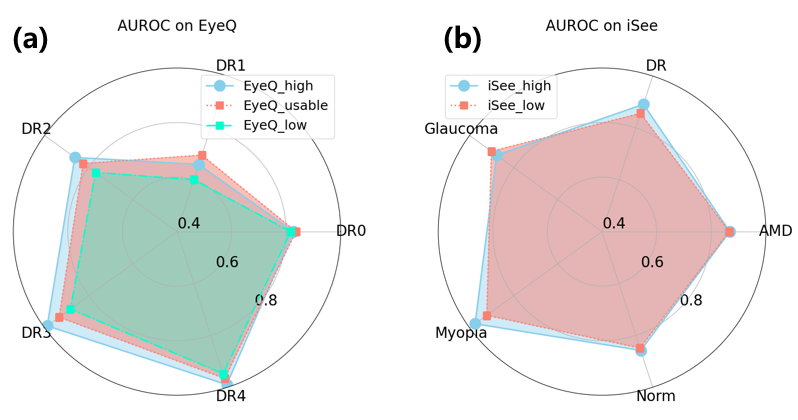}
        \par
    \end{centering}
\caption{RETFound fine-tuning on the high-quality EyeQ and then evaluated on low quality, usable quality, and high-quality EyeQ dataset. The AUROC of five classes is distributed in five DR gradings. For classes with lower proportions, RETFound is more significantly affected by image degradation.
}
\vskip -5pt
\label{fig:RETFound_rendar}
\end{figure}

\subsection{The Impact of Dataset Bias on RETFound}

To further investigate how each class in the dataset affects the performance of RETFound on high- and low-quality data subsets, we plotted the AUROC radar charts for RETFound on subsets of classes in EyeQ and iSee, as shown in Fig. \ref{fig:RETFound_rendar}. Detailed analysis and conclusions are presented below:

\textbf{EyeQ: }
From Fig. \ref{fig:RETFound_rendar} (a), we can see two interesting phenomena: one is the AUROC of DR-1 is lower than the average score, and the other one is that as the image quality decreases, the AUROC of DR-4 increases.
In our observation, 95\% of DR-1 cases were misclassified as either DR-0 or DR-2. This misgrading can be attributed to the larger proportion of DR-0 and DR-2 in the training dataset, as well as the similarity in pathological characteristics between DR-1 and DR-0/DR-2. Consequently, the AUROC for DR-1 is notably lower. Additionally, among the five DR grades, DR-0 is the least affected by image quality degradation, which may be attributed to DR-0 being the largest category in the EyeQ dataset.

\textbf{iSee: }Notably, from Fig. \ref{fig:RETFound_rendar} (b), it is peculiar that RETFound performs better in classifying glaucoma on low-quality subsets instead of high-quality subsets. This phenomenon may be attributed to two factors: firstly, the association of cataracts and glaucoma with age. We divided the iSee dataset into low-quality and high-quality based on the presence of cataracts. When evaluated on the cataract-dependent low-quality iSee, RETFound classifies more images as glaucoma. Secondly, the pathological characteristics of glaucoma commonly include optic disc excavation, characterized by an enlargement of the optic cup relative to the optic disc, thinning of the neuroretinal rim, and defects in the retinal nerve fiber layer. This phenomenon is exacerbated by the lack of global features in the cup-to-disc ratio within global images.

Moreover, the Norm and AMD categories are minimally affected by image quality degradation, due to Norm and AMD taking up most of iSee. Therefore, although the performance of RETFound is influenced by the quality of the images, balancing the proportions of each class in the dataset can mitigate these impacts.

\subsection{The Impact of Quality on Different Fine-tuning Techniques}

To explore how different fine-tuning methods affect the performance, we use three methods to fine-tune RETFound on iSee and EyeQ: The first one is modifying the classified layers of pre-trained RETFound, and then updating the whole model on the new dataset, which is called overall fine-tuning. The second one is freezing the encoding of RETFound, and then only updating the gradient of classified layers, which is called Linear Probe. The third one is to use the Tip-adapter to train the pre-trained RETFound, based on the pre-trained  CLIP model.

\begin{figure}[!t]
    \begin{centering}
        \includegraphics[width=0.9\linewidth]{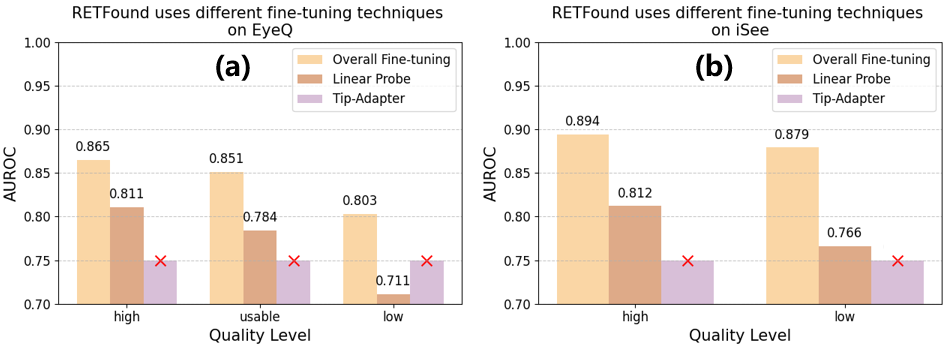}
        \par
    \end{centering}
\caption{The red cross (\textcolor{red}{×}) indicates the inability to compute AUROC. (a) shows that RETFound's AUROC is based on high-quality, usable-quality, and low-quality subsets of EyeQ, which is based on three fine-tuning techniques. (b) shows that the AUROC of RETFound on high-quality and low-quality subsets of iSee, which is based on three fine-tuning techniques. Among them, the overall fine-tuning approach demonstrates strong resistance to quality degradation. However, the Tip-Adapter method fails to calculate AUROC.
}
\vskip -5pt
\label{fig:fine-tune}
\end{figure}

From Fig. \ref{fig:fine-tune}, we can see that overall fine-tuning performs great, but the Tip-Adapter can not calculate AUROC. Two primary factors contribute to this phenomenon. Firstly, the presence of imbalanced datasets, where the 'norm' class dominates in iSee and the 'DR-0' class predominates in EyeQ, results in the model disproportionately focusing on these classes during training, while neglecting others. Secondly, the utilization of the CLIP model within the Tip-adapter introduces challenges. CLIP, a multimodal large-scale model, has been trained on a vast corpus of four billion text-image pairs, with limited representation of medical images. Consequently, the zero-shot classification performance of CLIP on medical images is suboptimal. Moreover, we also observed that the performance of RETFound, which uses the linear probe is inferior to the overall fine-tuning. The reason is that the linear probe failed to update the encoding part of RETFound. The encoding part retains most of the features from the pre-trained dataset, leading to inferior classification performance compared to the overall fine-tuning.





\section{Discussion and Conclusion}
To explore how the dataset quality impacts LFM, we conducted multiple experiments that evaluated RETFound, ResNet, and different fine-tuning techniques on distinct quality subsets of EyeQ and iSee. 

while RETFound exhibits high performance and generalizability in detecting fundus diseases and shows significant improvements in predicting systemic diseases, its performance is subject to the influence of dataset quality, with performance decreasing as image quality degrades

In conclusion, when encountering image quality degradation, LFM exhibits stronger robustness compared to the smaller model. Also, RETFound is susceptible to dataset bias, the lower the proportion of the category, the poorer the performance against image quality degradation of LFM in this category. For the fine-tuning techniques, overall fine-tuning RETFound can mitigate the impact of dataset quality on the model's performance.

%
%

\bibliographystyle{splncs04}
\bibliography{briefbib}

\begin{thebibliography}{10}
\providecommand{\url}[1]{\texttt{#1}}
\providecommand{\urlprefix}{URL }
\providecommand{\doi}[1]{https://doi.org/#1}

\bibitem{advances}
Cortes, C., Lawarence, N., Lee, D., Sugiyama, M., Garnett, R.: Advances in neural information processing systems 28. In: Proceedings of the 29th Annual Conference on Neural Information Processing Systems (2015)

\bibitem{cui2024scgpt}
Cui, H., Wang, C., Maan, H., Pang, K., Luo, F., Duan, N., Wang, B.: scgpt: toward building a foundation model for single-cell multi-omics using generative ai. Nature Methods pp. 1--11 (2024)

\bibitem{VIT}
Dosovitskiy, A., Beyer, L., Kolesnikov, A., Weissenborn, D., Zhai, X., Unterthiner, T., Dehghani, M., Minderer, M., Heigold, G., Gelly, S., et~al.: An image is worth 16x16 words: Transformers for image recognition at scale. arXiv preprint arXiv:2010.11929  (2020)

\bibitem{MCF}
Fu, H., Wang, B., Shen, J., Cui, S., Xu, Y., Liu, J., Shao, L.: Evaluation of retinal image quality assessment networks in different color-spaces. In: Medical Image Computing and Computer Assisted Intervention--MICCAI 2019: 22nd International Conference, Shenzhen, China, October 13--17, 2019, Proceedings, Part I 22. pp. 48--56. Springer (2019)

\bibitem{he2023accuracy}
He, S., Bao, R., Li, J., Grant, P.E., Ou, Y.: Accuracy of segment-anything model (sam) in medical image segmentation tasks. arXiv preprint arXiv:2304.09324  (2023)

\bibitem{huang2024segment}
Huang, Y., Yang, X., Liu, L., Zhou, H., Chang, A., Zhou, X., Chen, R., Yu, J., Chen, J., Chen, C., et~al.: Segment anything model for medical images? Medical Image Analysis  \textbf{92},  103061 (2024)

\bibitem{SAM}
Kirillov, A., Mintun, E., Ravi, N., Mao, H., Rolland, C., Gustafson, L., Xiao, T., Whitehead, S., Berg, A.C., Lo, W.Y., et~al.: Segment anything. arXiv preprint arXiv:2304.02643  (2023)

\bibitem{koonce2021resnet}
Koonce, B., Koonce, B.: Resnet 50. Convolutional Neural Networks with Swift for Tensorflow: Image Recognition and Dataset Categorization pp. 63--72 (2021)

\bibitem{SAME}
Li, H., Lin, Z., Qiu, Z., Li, Z., Niu, K., Guo, N., Fu, H., Hu, Y., Liu, J.: Enhancing and adapting in the clinic: Source-free unsupervised domain adaptation for medical image enhancement. IEEE Transactions on Medical Imaging  (2023)

\bibitem{SCR}
Li, H., Liu, H., Fu, H., Shu, H., Zhao, Y., Luo, X., Hu, Y., Liu, J.: Structure-consistent restoration network for cataract fundus image enhancement. In: International Conference on Medical Image Computing and Computer-Assisted Intervention. pp. 487--496. Springer (2022)

\bibitem{li2023artificial}
Li, X., Zhang, L., Wu, Z., Liu, Z., Zhao, L., Yuan, Y., Liu, J., Li, G., Zhu, D., Yan, P., et~al.: Artificial general intelligence for medical imaging. arXiv preprint arXiv:2306.05480  (2023)

\bibitem{lin2023domain}
Lin, Y., Li, H., Liu, H., Shu, H., Li, Z., Hu, Y., Liu, J.: Domain adaptative retinal image quality assessment with knowledge distillation using competitive teacher-student network. In: 2023 IEEE 20th International Symposium on Biomedical Imaging (ISBI). pp.~1--5. IEEE (2023)

\bibitem{clip}
Radford, A., Kim, J.W., Hallacy, C., Ramesh, A., Goh, G., Agarwal, S., Sastry, G., Askell, A., Mishkin, P., Clark, J., et~al.: Learning transferable visual models from natural language supervision. In: International conference on machine learning. pp. 8748--8763. PMLR (2021)

\bibitem{attention}
Vaswani, A., Shazeer, N., Parmar, N., Uszkoreit, J., Jones, L., Gomez, A.N., Kaiser, {\L}., Polosukhin, I.: Attention is all you need. Advances in neural information processing systems  \textbf{30} (2017)

\bibitem{wu2023medical}
Wu, J., Fu, R., Fang, H., Liu, Y., Wang, Z., Xu, Y., Jin, Y., Arbel, T.: Medical sam adapter: Adapting segment anything model for medical image segmentation. arXiv preprint arXiv:2304.12620  (2023)

\bibitem{tip}
Zhang, R., Zhang, W., Fang, R., Gao, P., Li, K., Dai, J., Qiao, Y., Li, H.: Tip-adapter: Training-free adaption of clip for few-shot classification. In: European Conference on Computer Vision. pp. 493--510. Springer (2022)

\bibitem{RETFound}
Zhou, Y., Chia, M.A., Wagner, S.K., Ayhan, M.S., Williamson, D.J., Struyven, R.R., Liu, T., Xu, M., Lozano, M.G., Woodward-Court, P., et~al.: A foundation model for generalizable disease detection from retinal images. Nature  \textbf{622}(7981),  156--163 (2023)

\end{thebibliography}


%




\end{document}